\documentclass{aa}

\usepackage{natbib}
\bibpunct{(}{)}{;}{a}{}{,} 
\usepackage[colorlinks=true,allcolors=blue]{hyperref}%
\usepackage{amsmath, amssymb}
\usepackage{physics}         
\usepackage{mathtools}       
\usepackage{graphicx}
\usepackage{txfonts}

\begin{document}

   \title{Beyond the two-point correlation: constraining primordial non-gaussianity with density perturbation moments}


   \author{E. Vislosky\inst{1,2}\thanks{\email{ethanvis322@gmail.com}}
       \and Z. Brown\inst{1,3}
       \and R. Demina\inst{1}
       \and E. Chaussidon\inst{4}}

   \institute{Department of Physics and Astronomy, University of Rochester 500 Joseph C. Wilson Boulevard, Rochester, NY 14627, USA
   \and Center for Astrophysics--Harvard \& Smithsonian, 60 Garden Street, Cambridge, MA 02138, USA
   \and Department of Physics, Kansas State University, 116 Cardwell Hall, Manhattan, KS 66506, USA
   \and Lawrence Berkeley National Laboratory, 1 Cyclotron Road, Berkeley, CA 94720, USA}

   \date{Received XX XX, XXXX; accepted YY YY, YYYY}

 
  \abstract
   {Constraining primordial non-Gaussianities (PNGs) in the large-scale cosmic structure (LSS) is an important step in understanding properties of the early universe, specifically in distinguishing between different inflationary models.  Measuring PNG relies on evaluating the scale-dependent correlations in the density field. New summary statistics beyond the two- and three-point correlation functions in configuration space and their Fourier-space counterparts, the power- and bi-spectrum may provide increased sensitivity.}
   {We introduce a new method for extracting the PNG signal imprinted on the LSS by using the first three Gaussian moments of the normalized correlation in density perturbations, evaluated at varying distance scales. We aim to assess this method's sensitivity to local PNG, parameterized by $f_{\text{NL}}$.}
   {We perform spherical convolutions at a range of scales on dark matter halo simulations to measure the scale-dependent correlations in the density field. From these, we compute the first three moments and compare them to a model expectation vector, parameterized to the second power in $f_{\text{NL}}$.}
   {Our method provides about 21\% improvement in sensitivity to $f_{\text{NL}}$ with respect to using the two point correlation function alone. Notably, we find that the second moment alone carries nearly as much constraining power as the mean, highlighting the potential of higher-order statistics. Given its simplicity and efficiency, this framework is well-suited for application to current and upcoming large-scale surveys such as DESI.}
   {}

   \keywords{
    cosmology: inflation --
    cosmology: large-scale structure of Universe --
    methods: statistical --
    early Universe
   }

   \maketitle{}
%
%
\section{Introduction}

    It is now widely accepted within the cosmological community that there existed an epoch of rapid spacetime expansion in the early Universe, referred to as inflation. This inflationary epoch is characterized by exponential growth of the scale factor in such a way that neatly explains the isotropy and homogeneity we observe in the large-scale structure (LSS) today. The driving mechanism for inflation is believed to be scalar field oscillations \citep{Guth81,Linde82,Albrecht1982}. Despite the remarkable theoretical success of inflation in its ability to solve the horizon and flatness problems \citep{Guth81}, among others, its detailed characteristics are not well constrained. Thus, we turn to observational probes.

    In the simplest inflationary scenario, involving a single scalar field, over-densities in the cosmic matter-density profile are expected to follow a Gaussian distribution with mean zero \citep{Maldacena03}. Conversely, multi-field models predict varying degrees of deviation from Gaussianity \citep{Creminelli11}. To distinguish between these models and refine our understanding of the early Universe, we rely on measurements of local primordial non-Gaussianity (PNG).

    The primordial gravitational potential for a general scalar field model can be expressed as: 
    \begin{equation}\label{eqn:fNL}
        \Phi_{\text{NG}} = \phi +f_{\text{NL}}(\phi^2 - \langle \phi \rangle^2), 
    \end{equation} 
    where $\phi$ represents the potential of a Gaussian random field, and $f_{\text{NL}}$ (this is really $f_{\text{NL}}^{\text{loc}}$, the local PNG parameter, but throughout we will refer to it just as $f_{\text{NL}}$) quantifies the amplitude of the quadratic deviation in the potential \citep{Komatsu01, Gangui94}. A non-zero value of $f_{\text{NL}}$ indicates deviations from the single-field model. Consequently, constraining $f_{\text{NL}}$ is a key step in testing inflationary models.

    Significant efforts have been made to measure $f_{\text{NL}}$. One of the primary methods involves analyzing anisotropies in the cosmic microwave background (CMB), with the Planck bispectrum analysis providing the tightest constraint to date: $f_{\text{NL}}=0.9\pm5.1$ (68\% confidence limits) \citep{Planck19}. While this has ruled out some highly non-Gaussian multi-field models, achieving greater precision is necessary to further narrow down the model parameter space. Moreover, CMB-based analyses are fundamentally limited by cosmic variance \citep[e.g.,][]{Azabajian16}.

    Alternative approaches to probing PNGs involve the LSS, offering opportunities for cross-validation and potentially greater sensitivity to $f_{\text{NL}}$. These methods analyze the distribution of galaxy clusters and dark matter halos through the power spectrum and bispectrum, among higher-order statistics \citep[see, e.g.,][]{Ross12, Castorina19, mueller21, cagliari25}, which are Fourier transforms of the configuration-space two-point, three-point, and higher-order correlation functions (2pcf, 3pcf, and \textit{n}pcf, respectively)\citep{Brown24}. More recently, it has also been shown that field-level inference can be applied as a comprehensive alternative to summary statistics in probing PNG \citep{andrews24}. Current LSS-based constraints on $f_{\text{NL}}$ find $\sigma_{f_{\text{NL}}} \approx 9$ \citep{chaussidon2024}. However, these analyses require careful characterization of nonlinearities induced by structure formation, which must be disentangled from those originating from PNG, making tighter constraints increasingly difficult.
    
    Despite these efforts, the sensitivity required to differentiate even the simplest inflationary models remains elusive. For instance, the slow-roll scenario \citep{Guth81, Linde82} predicts $f_{\text{NL}} \sim \mathcal{O}(0.01)$ \citep{Maldacena03}. Achieving such precision demands novel methodologies and advancements in observational capabilities.
    
    In this paper, we introduce a new method for probing PNG using the large-scale structure. Since PNG introduces scale-dependent correlations in the density field—particularly at large scales—traditional two-point statistics capture only part of the signal. By contrast, higher-order moments of the density perturbation field contain information about the width and skewness of the distribution, which are also sensitive to PNG. We propose a method that computes the first three Gaussian moments of a normalized two-point estimator across multiple scales and uses these to constrain $f_{\text{NL}}$.

    We test this method on three sets of dark matter (DM) halo simulations, each seeded with $f_{\mathrm{NL}} = -25,\ 0,\ \text{and}\ 25$, collectively referred to as the FastPM-L1 simulations \citep{chaussidon23}. Each set contains 125 cubic boxes of side length $L = 1h^{-1}\,\mathrm{Gpc}$.\footnote{The reduced Hubble constant is defined as $h=H_0 / (100\ \mathrm{km/s/Mpc})$, where $H_0$ is the present-day Hubble constant.} Using the \texttt{ConKer} algorithm \citep{Brown22}, we estimate the correlation in matter density perturbations at various scales. From these distributions, we compute an observation vector $\hat{\mu}$, consisting of the first three moments of the density perturbation distributions (see Sec.~\ref{sec:den_moments}). Parameterizing these vectors as functions of $f_{\text{NL}}$ yields the expectation vector $\tilde{\mu}(f_{\text{NL}})$. We then construct a $\chi^2$ test statistic and use a Markov chain Monte Carlo (MCMC) sampler to evaluate the method’s sensitivity on an independent $f_{\mathrm{NL}}=12$ FastPM-L1 simulation set, comparing these results to that of the 2pcf. As we will show, the second moment alone provides nearly comparable constraining power to the 2pcf, underscoring its value as a standalone observable. Our aim is not to report absolute constraints on $f_{\text{NL}}$, but rather to assess the relative improvement in sensitivity offered by the moments method and characterize its inherent predictive capabilities.

    This moments-based approach was previously explored by \cite{mao2014}, who showed that moments of the density field could distinguish between Gaussian and non-Gaussian initial conditions. Our study builds on their insights by applying the method to simulations with $f_{\text{NL}}$ values closer to current observational limits. We formalize a statistical model, improve uncertainty quantification using a larger ensemble of simulations, and demonstrate enhanced sensitivity to PNG—highlighting the method’s potential for modern cosmological analyses.


\section{Method}

\subsection{Theoretical considerations}

    By applying the Poisson equation to the primordial gravitational potential in Eq.~\eqref{eqn:fNL}, and only considering high peaks in the density distribution--where halos form \citep{dalal2008}--the matter overdensity traced by halos in the presence of non-Gaussianity becomes
    \begin{equation}\label{eqn:NGden}
        \delta_{h}\approx[b_1+b_{\phi}\alpha^{-1} f_{\text{NL}}]\delta_{\text{NG}}.
    \end{equation}
    Here, $\delta_{NG}$ is the density field sourced by the potential $\Phi_{\text{NG}}$ (Eq.~\eqref{eqn:fNL}),  $b_1$ is a linear bias that relates the overdensity of halos to that of matter, and  $b_{\phi }$ is the PNG bias, degenerate with $f_{\mathrm{NL}}$. A typical parameterization of $b_{\phi }$ is:
    \begin{equation}
        b_{\phi }=2\delta_c (b_{1}-p),
    \label{eq:bphi}
    \end{equation}
    where $\delta_c=1.69$ is the critical density threshold for spherical collapse, and $p$ describes halo formation in primordial overdensities \citep{barreira2020impact}. We assume the universality relation $p=1$, consistent with our halo mass cut (see Sec.~\ref{sec:disc:mass} and \citealt{adame2024}). 
    
    The scale-dependent factor $\alpha$ is a function of the wave-number $k$ associated with density perturbations on scales $s \sim k^{-1}$ and the redshift $z$:
    \begin{equation}
        \alpha(k,z)=\frac{2c^2 D(z)}{3 H_0^2 \Omega_{m} }  \frac{g(z_{\mathrm{rad}})}{g(0)} T(k) k^2,
    \label{eq:alpha_k}
    \end{equation}
    where $c$ is the speed of light,  $\Omega_{m}$ is the present day relative matter density, and $D(z)$ is the growth factor normalized to 1 at $z=0$. The factor $g(z_{\mathrm{rad}}) / g(0)$ takes into account the difference of this  normalization with respect to the early time matter-dominated epoch. The transfer function $T(k)$ is close to unity for scales $s>L_0=16(\Omega_{m,0} h^2)^{-1}\approx 3h^{-1}\,\mathrm{Mpc}$, which are typically considered in LSS clustering analyses. Thus, the PNG-generated density perturbations depend quadratically on the scale $s$.
    
    To probe the imprint of these scale-dependent, non-Gaussian features on the LSS, we compute the correlation of the matter perturbation field:
    \begin{equation}\label{eqn:dcp}                
        \eta(\vec{x},s)\equiv\delta(\vec{x})\delta(\vec{x}+s),
    \end{equation}
    where bold variables represent three-dimensional vectors. Note that in this notation, the standard two-point correlation function corresponds to the spatial average of $\eta$—that is, the average over all pairs of points separated by a distance $s$:
    \begin{equation}\label{eqn:2pcf}
        \xi_2(s)\equiv\langle\eta(\vec{x},s)\rangle.
    \end{equation}

\subsection{Evaluation of matter density field}\label{sec:eval_field}

    \begin{figure}[htbp]
    \includegraphics[width=9cm]{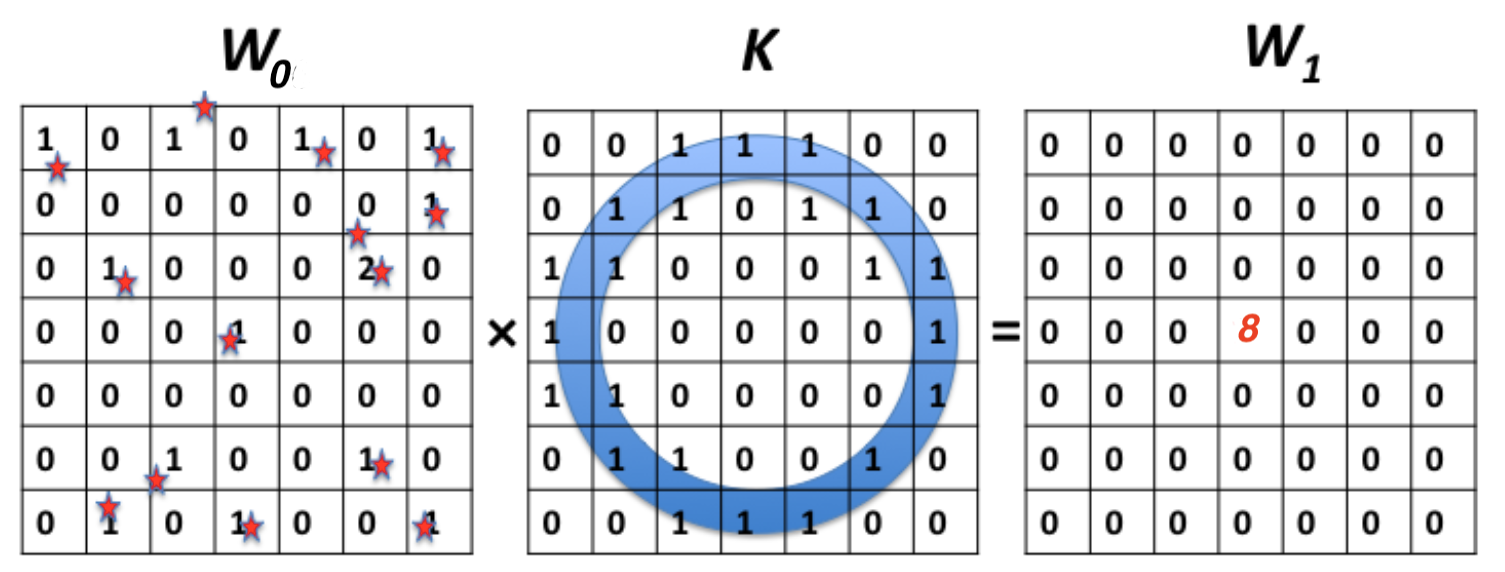}
    \caption{Two-dimensional schematic of how the \texttt{ConKer} algorithm evaluates a density field $W_0$ and $W_1$. This shows the convolution of the matter-density variation field, $W_0=D-R$ (left), with a spherical shell kernel $K$ of radius \textit{s} and width $\Delta s$. The resulting product is the density histogram $W_1$ (right). In this example, the inner product is 8, which is placed in the center cell of the $W_1$ grid. The kernel center iterates over each cell, performing this convolution at each scale $s$.}
    \label{fig:convolve}
    \end{figure}

    \noindent
    The matter density field in simulations and real data is evaluated on a discrete grid. In this setting, the density contrast product from Eq.~\eqref{eqn:dcp} becomes a discrete convolution:
    \begin{equation}\label{eqn:dcpD}
        \eta(\vec{x},s)=\delta(\vec{x})[\delta(\vec{x})***K(s)],
    \end{equation}
    where $K(s)$ is a spherical shell kernel of radius $s$ and thickness $\Delta s$, and $\ast \ast \ast$ denotes a 3D discrete convolution.

    We compute this quantity using the \texttt{ConKer} algorithm \citep{Brown22}, which operates on the density contrast grid
    \begin{equation}\label{eqn:W0}
        \noindent
        W_0(X,Y,Z)\equiv D(X,Y,Z) - R(X,Y,Z)
    \end{equation}
    \noindent
    where $D(X,Y,Z)$ and $R(X,Y,Z)$ represent the data and random tracer counts in the cell $(X,Y,Z)$, respectively, and the full catalog is divided into $N$ such cells.
    
    Next, we convolve $W_0$ with spherical kernels across a range of scales $s = 55h^{-1}\,\mathrm{Mpc}-395h^{-1}\,$Mpc, in bins of width $\Delta s=20h^{-1}\,$Mpc. (For our actual test cases, the maximum scale $s$ we use may vary; see Sec.~\ref{sec:simulations}, Appendix~\ref{app:sdep} for justification). This procedure yields a histogram grid at each scale:
    \begin{equation}\label{eqn:W1}
        \noindent
        W_1(X,Y,Z;s)\equiv W_0(X,Y,Z) \ast \ast \ast K(s), 
    \end{equation}
    \noindent
    as illustrated in Fig.~\ref{fig:convolve}.

    We apply the same procedure to the random catalog to compute
    \begin{equation}\label{eqn:B1}
        \noindent
        B_1(X,Y,Z;s)\equiv B_0(X,Y,Z) \ast\ast\ast K(s),
    \end{equation}
    \noindent
    where $B_0$ is just the random catalog tracer counts $R(X,Y,Z)$.
    \indent
    In this notation $\delta(\vec{x})\sim W_0(X,Y,Z)$ and $\delta(\vec{x})***K(s)\sim W_1(X,Y,Z;s)$, so the discrete density contrast product from Eq.~\eqref{eqn:dcpD} becomes:
    \begin{equation}\label{eqn:dcpW}
        \eta{\vec(X,Y,Z;s)=W_0(X,Y,Z)\cdot W_1(X,Y,Z;s)}.
    \end{equation}
    
    To construct a scalar field suitable for moment analysis, we flatten $\eta(X,Y,Z;s)$ into a one-dimensional array and normalize it using the spatial average random field product $\langle \eta_R(s)\rangle$, yielding:
    \begin{equation}\label{eqn:dist}
        \Gamma_n(s)\equiv\frac{\eta(X,Y,Z;s)}{\langle \eta_R(s)\rangle}\frac{N_R^2}{N_D^2},
    \end{equation}
    \noindent
    where $n$ indexes over the flattened spatial grid, and $N_D$ and $N_R$ are the total counts of data and random tracers, respectively. This normalization mitigates boundary artifacts and ensures statistical consistency across simulations.

\begin{figure}[htbp]
    \includegraphics[width=9cm]{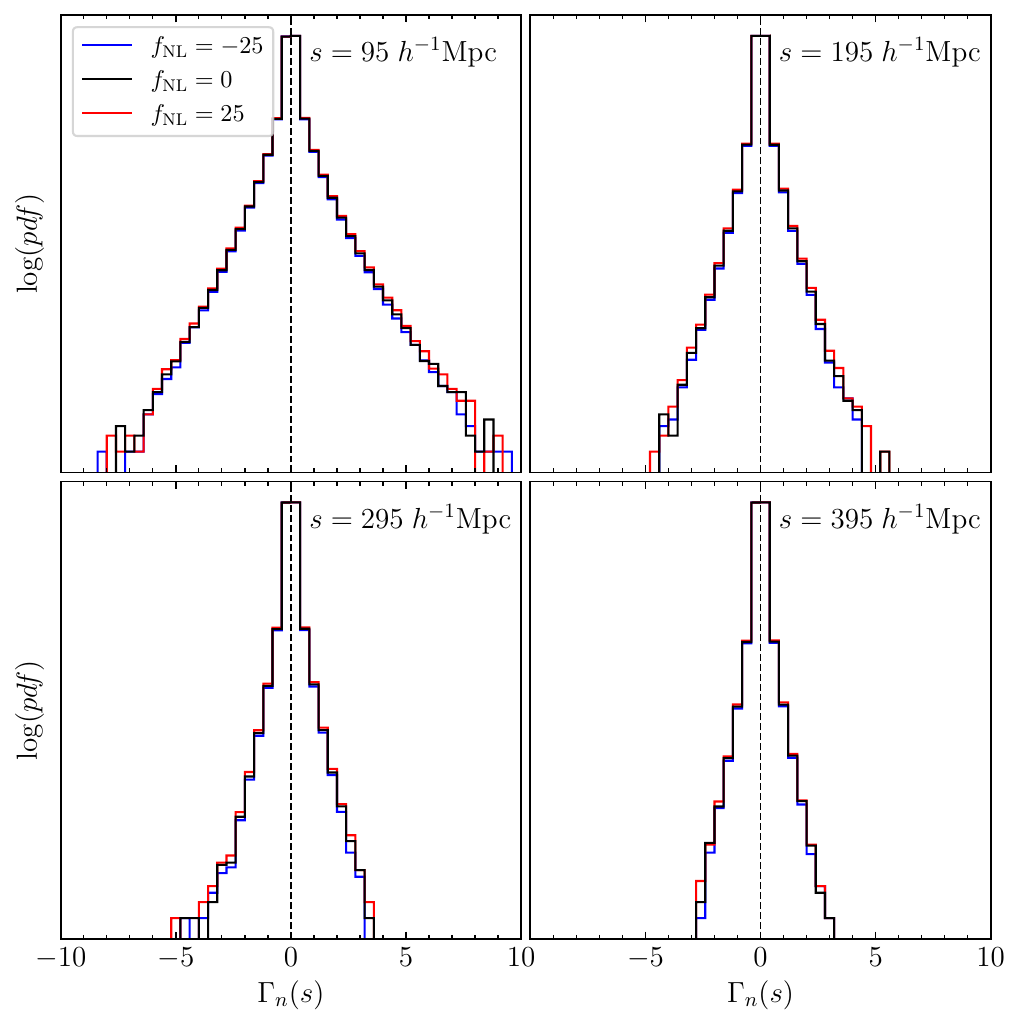}
    \caption{Distributions of the flattened density perturbation estimator $\Gamma_n(s)$, averaged over all 125 FastPM-L1 simulations, $\langle\Gamma_n(s)\rangle_m$, shown at $s=95,195,295$, and $395h^{-1}\,$Mpc for the $f_{\text{NL}}=$-25 (blue), 0 (black), and 25 (red) cases.
    }
    \label{fig:distributions}      
    \end{figure}

    The resulting $\Gamma_n(s)$ distributions encode scale-dependent deviations from Gaussianity. As seen in Fig.~\ref{fig:distributions}, these deviations manifest themselves as variations in the shape (particularly the width and skewness) of the distributions at different scales, characterized by the value of $f_{\text{NL}}$. This behavior underscores the sensitivity of $\Gamma_n(s)$ to PNG. This motivates our next step: quantifying these deviations by computing the Gaussian moments (referred to as moments for brevity) of the distributions, which provide a compact and statistically robust characterization of the underlying PNG.
    
\subsection{Density perturbation moments}\label{sec:den_moments}

    To probe non-Gaussian features in the matter density field, we compute the first three moments of the flattened distributions $\Gamma_n(s)$. This is motivated by the fact that information about a distribution's shape is fully encoded in a series expansion of its moments. The first moment, $\mu_1(s)$ is equivalent to the mean (analagous to the 2pcf, see Sec.~\ref{sec:mom}), the second, $\mu_2(s)$ is the variance, and the third $\mu_3(s)$ is the skewness of the distribution:
    \begin{align}
        \mu_1(s) &= \frac{1}{N}\sum_{n} \Gamma_n(s) \label{eqn:mu1} \\
        \mu_2(s) &= \frac{1}{N}\sum_{n} \left[\Gamma_n(s)-\mu_1(s)\right]^2 \label{eqn:mu2} \\
        \mu_3(s) &= \frac{1}{N}\sum_{n} \left[\Gamma_n(s)-\mu_1(s)\right]^3, \label{eqn:mu3}
    \end{align}
    \noindent
    where the summations are taken over the \textit{N} spatial grid-points in the density perturbation-estimator. See Appendix~\ref{app:momphys} for a deeper, physical interpretation of the density moments.

    These three moments cumulatively characterize the central tendency, spread, and asymmetry of the distribution. Since a function is fully encoded in its infinite series of moments, even a truncated expansion can efficiently capture deviations from Gaussianity.
    
\begin{figure*}
    \centerline{\includegraphics[width=18cm]{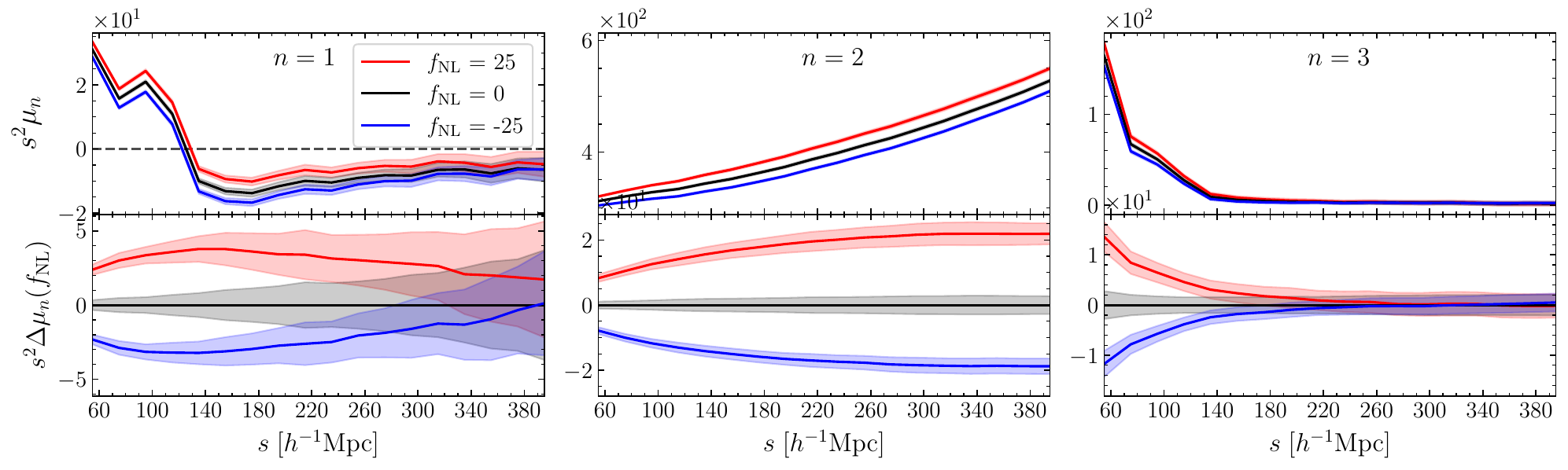}}
    \caption{The moments $\mu_1$, $\mu_2$, and $\mu_3$ (left to right) computed from the average density fields of the 125 FastPM-L1 simulations, shown for $f_{\text{NL}}=$-25 (blue), 0 (black), and 25 (red). Top panels show raw values; bottom panels show residuals relative to the $f_{\text{NL}} = 0$ case. Shaded regions represent the standard error across simulations. All moments are scaled by $s^2$ to highlight large-scale structure.}
    \label{fig:moments}      
    \end{figure*}

    As illustrated in Fig.~\ref{fig:moments}, these functions have a noticeable dependence on the $f_{\text{NL}}$ parameter, with the moments from the positive (negative) $f_{\text{NL}}$ case having greater (lesser) magnitudes than those for the fiducial $f_{\text{NL}}=0$ case. More specifically, since $W_0\sim\delta$, and $\delta$ varies linearly with $f_{\text{NL}}$ from Eq.~\eqref{eqn:NGden}, we expect the density grids $W$ to exhibit this same dependence. Therefore, given that our density perturbation-estimator $\Gamma_n(s)$ includes a product of these density fields in the form of Eq.~\eqref{eqn:dcpW}, we expect the moments to exhibit a quadratic dependence in $f_{\text{NL}}$ (see Fig.~\ref{fig:quadfit} and Eq.~\eqref{eqn:exp_vec}). So, in constructing a combined vector of these moments we obtain a tool sensitive to PNG. Note that this dependence is only complete for $\mu_1$, higher moments include higher powers of $\Gamma_n(s)$ and thus higher order terms in $f_{\text{NL}}$ ($f_{\text{NL}}^4$ and $f_{\text{NL}}^6$ for the 2nd and 3rd moments, respectively; see Sec.~\ref{sec:discuss} for further comments). In practice however, we find that quadratic parametrization is sufficient for the range of scales considered in this analysis.
    
    We concatenate the three moments into a single observation vector:
    \begin{equation}\label{eqn:obs_vec}
        \hat{\mu}_i = \left[\mu_1(s),\;\mu_2(s),\;\mu_3(s)\right],
    \end{equation}\noindent
    where $i$ indexes over $s$-bins. For 18 bins per moment, this yields a total of $3\times18=54$ elements in the vector. As we demonstrated, each individual moment follows a roughly quadratic dependence on $f_{\text{NL}}$, so we expect the full observation vector to exhibit similar behavior. Accordingly, for a catalog with arbitrarily seeded PNG, we model the expectation vector as:
    \begin{equation}\label{eqn:exp_vec}
        \tilde{\mu}_i(f_{\text{NL}}) = \mathcal{A}_if_{\text{NL}}^2 + \mathcal{B}_if_{\text{NL}} + \mathcal{C}_i,
    \end{equation}
    \noindent
    where $\mathcal{C}_i = \hat{\mu}_i(f_{\text{NL}} = 0)$. The coefficients $\mathcal{A}_i$ and $\mathcal{B}_i$ encode the expected scale-dependent response of each moment to PNG, and are determined from fits to multiple simulations with known values of $f_{\text{NL}}$.

\section{Application to simulations}

\subsection{Simulations}\label{sec:simulations}

    Approximately 80\% of the matter in the Universe is believed to be composed of dark matter (DM), which forms massive, gravitationally bound structures called halos--within which galaxies form. By simulating a Universe composed entirely of DM particles, we can effectively emulate its large-scale structure (see Eq.~\eqref{eqn:NGden}).
    
    For this study, we use a suite of four approximated particle-mesh \textit{N}-body simulations generated with \texttt{FastPM} \citep{feng16}, each seeded with local PNG values of $f_{\text{NL}} = -25,\;0,\;12,$ and $25$ \citep[see run-knl-3 in][]{chaussidon23}. The non-Gaussian initial conditions (via Eq.~\eqref{eqn:fNL}) are imposed at redshift $z=19$, and the simulations are evolved to $z=1.5$, covering a volume of $(L=5.52h^{-1}\,\mathrm{Gpc})^3$ with particle masses of $\sim 10^{11}h^{-1}M_{\odot}$.
    
    Each simulation volume is divided into $N_m = 125$ non-overlapping sub-boxes of $(L=1h^{-1}\,\mathrm{Gpc})^3$, which we refer to as the FastPM-L1 simulations (the remaining $0.52h^{-1}\,\mathrm{Gpc}$ is discarded). This subdivision into "mini-universes" is justified by the cosmological assumption of homogeneity and enables us to compute a covariance matrix (see Eq.~\eqref{eqn:cov}) from 125 independent mock realizations per $f_{\mathrm{NL}}$(see Sect.~\ref{sec:disc:limits} for caveats).
    
    To ensure a consistent tracer population and preserve the universality relation ($p=1$; see \citealt{adame2024} and Sect.~\ref{sec:disc:mass}), we apply a minimum halo mass cut of $M_h > 5 \times 10^{12}h^{-1}M_{\odot}$, yielding $N_D \approx 5.5 \times 10^5$ halos per box. A corresponding random catalog is generated by populating a $(L = 1h^{-1}\,\mathrm{Gpc})^3$ box with $N_R = 10N_D$ randomly distributed objects.

    We use the $f_{\mathrm{NL}} = -25,\;0,$ and $25$ simulations to build our model, reserving the $f_{\text{NL}}=12$ simulations for validation. From each box, we evaluate the scale-dependent density perturbation field $\Gamma_n(s)$ as described in Sect.~\ref{sec:eval_field}, and compute the average field $\langle\Gamma_n(s)\rangle_m$ across all 125 boxes for each $f_{\mathrm{NL}}$ value. These simulation-averaged distributions form the basis for our moment vectors (see Figs.~\ref{fig:distributions} and~\ref{fig:moments}). Unless otherwise noted, all references to $\Gamma_n(s)$ refer to these averaged fields. Note: a distinct random catalog is generated per sub-box to preserve statistical independence.
    
\begin{figure}[htbp]\centering 
    \includegraphics[width=9cm]{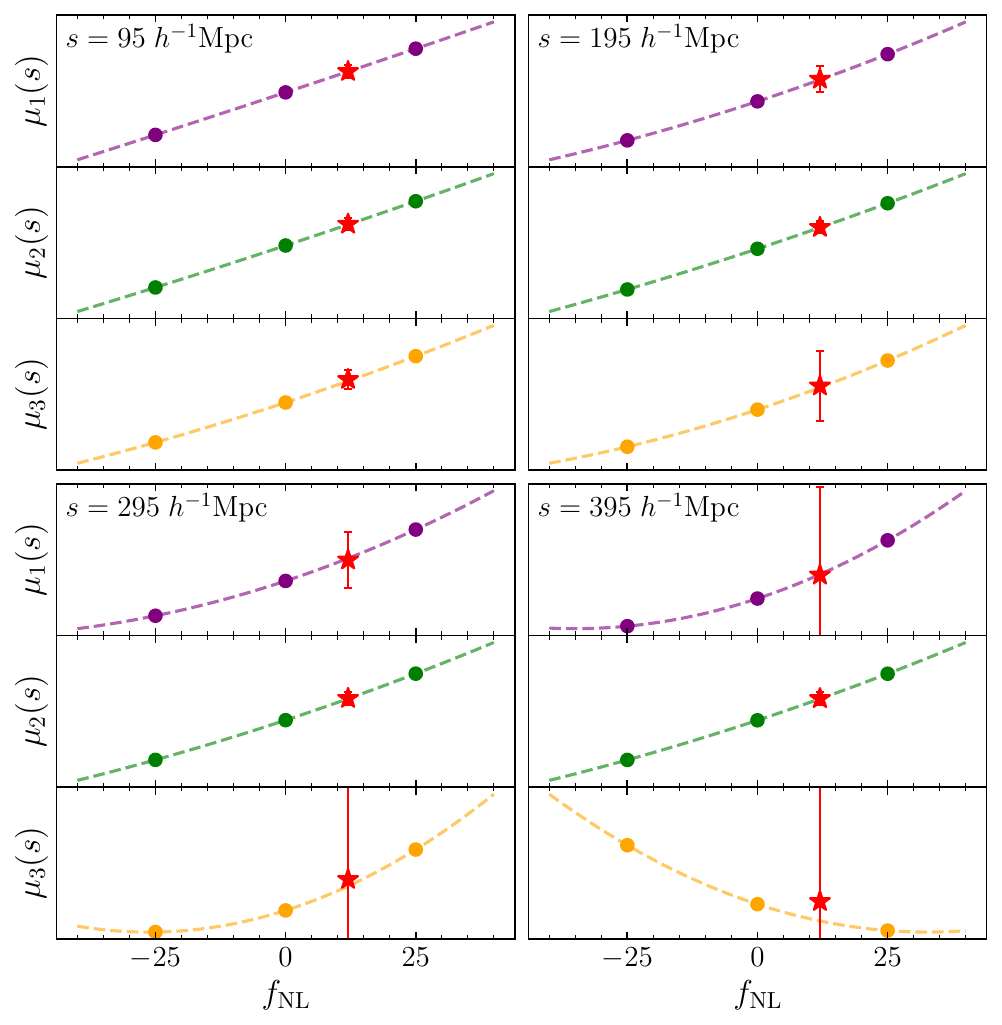}
    \caption{First (purple), second (green), and third (orange) moments plotted as a function of $f_{\text{NL}}$ for the same \textit{s}-bins as in Fig.~\ref{fig:distributions}. Each  point represents the average moment computed from the 125 FastPM-L1 simulations at $f_{\text{NL}}=-25,0,$ and $25$, comprising the observation vector $\hat{\mu}_i$ (with each moment plotted separately). Dashed lines indicate the quadratic fit--i.e., the expectation vector $\tilde{\mu}_i(f_{\text{NL}})$. The $f_{\text{NL}}=12$ test case is shown as a red star, with error bars representing the standard error on $\mu$ across the 125 simulations.}
    \label{fig:quadfit} 
    \end{figure}

    From these averaged fields, we compute the moment vectors $\hat{\mu}_i(f_{\mathrm{NL}})$ using Eq.~\eqref{eqn:obs_vec}, yielding three sets of observations at $f_{\mathrm{NL}} = -25,\;0,$ and $25$. We then fit these with a quadratic model (Eq.~\eqref{eqn:exp_vec}) to extract the coefficients:
    \begin{align}
        \mathcal{A}_i &= \left[\hat{\mu}_i(25)+\hat{\mu}_i(-25)-2\hat{\mu}_i(0)\right]/1250 \label{eqn:Acoeff}, \\
        \mathcal{B}_i &= \left[\hat{\mu}_i(25)-\hat{\mu}_i(-25)\right]/50, \label{eqn:Bcoeff} \\
        \mathcal{C}_i &= \hat{\mu}_i(0). \label{eqn:Ccoeff}
    \end{align}
    \noindent
    These yield the expectation vector $\tilde{\mu}_i(f_{\text{NL}})$ used throughout the analysis. Fig.~\ref{fig:quadfit} shows the fits at selected scales.

\subsection{Statistical framework}\label{sec:stat}

    To evaluate the robustness of our model and assess its sensitivity to PNG, we test it against an independent catalog with known $f_{\text{NL}}$. First, to characterize uncertainties in our observation vector, we derive a covariance matrix from the $N_m=$125 $f_{\text{NL}}=0$ simulations:
    \begin{equation}\label{eqn:cov}
        C_{ij}=\frac{1}{N_m-1}\sum_{m}\left[\hat{\mu}_{m,i}-\langle\hat{\mu}\rangle_i\right]\left[\hat{\mu}_{m,j}-\langle\hat{\mu}\rangle_j\right],
    \end{equation}
    \noindent
    where $\hat{\mu}_{m,i}$ is the value of the $i$-th bin from the $m$-th simulation, and $\langle\hat{\mu}\rangle_i$ is the mean of that bin over all $N_m$ realizations. The index $j$ iterates over each bin independently of $i$, corresponding to the second dimension of the covariance matrix. The covariance matrix structure is shown in Fig.~\ref{fig:cov}, computed from the fiducial $f_{\mathrm{NL}}=0$ FastPM-L1 set following standard practice (see Appendix~\ref{app:cov}).
    
    Next, we construct a test statistic to quantify the agreement between the observed vector $\hat{\mu}_i$ and the expected vector $\tilde{\mu}_i(f_{\text{NL}})$. Assuming a Gaussian likelihood, the $\chi^2$ statistic is given by:
    \begin{equation}
        \chi^2(f_{\text{NL}})=\sum_{ij}\left[\hat{\mu}_i-\tilde{\mu}_i(f_{\text{NL}})\right]^TC^{-1}_{ij}\left[\hat{\mu}_j-\tilde{\mu}_j(f_{\text{NL}})\right].
    \end{equation}
    The best-fit $f_{\mathrm{NL}}$ is then obtained by minimizing $\chi^2$ with respect to $f_{\mathrm{NL}}$ for the given observation vector.
    
    In Sec.~\ref{sec:sensitivity}, we apply this formalism to the $f_{\text{NL}} = 12$ FastPM-L1 simulation set and compare the performance of the full moment vector against its individual components.

    \begin{figure}[htbp]\centering
    \includegraphics[width=9cm]{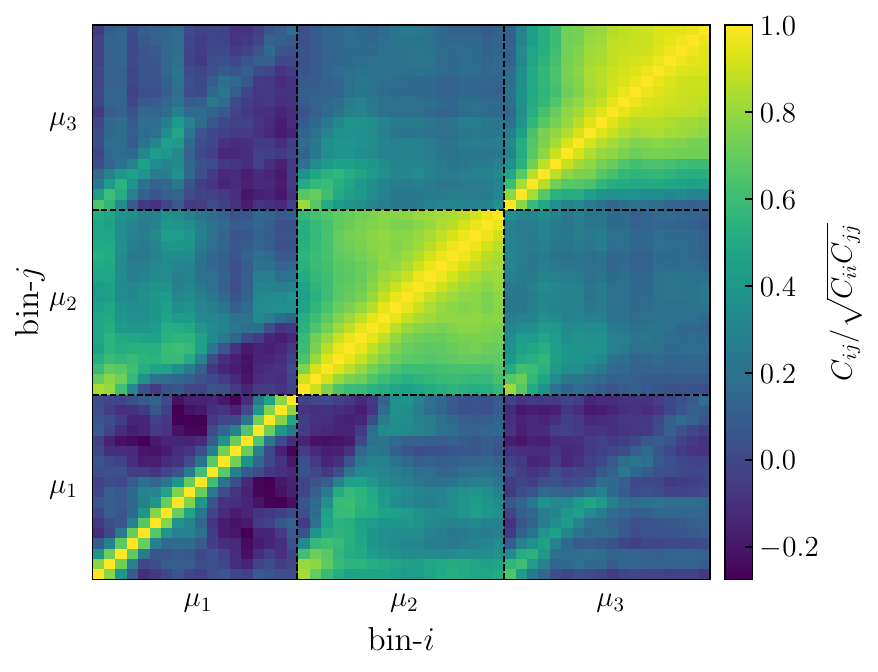}
    \caption{Correlation matrix ($C_{ij}/\sqrt{C_{ii}C_{jj}}$) computed from the 125 $f_{\text{NL}}=0$ simulations using Eq.~\eqref{eqn:cov}. The matrix here is derived using the full observation vector out to $s_{\mathrm{max}}=395h^{-1}\,$Mpc (54 total bins, 18 per moment). Dashed lines separate the blocks corresponding to different moments in the observation vector $\hat{\mu}_i$.}
    \label{fig:cov}
    \end{figure}

\section{Results: sensitivity to PNG}\label{sec:sensitivity}

    \subsection{$f_{\mathrm{NL}}=12$ test case}\label{sec:test_case}

        To evaluate the sensitivity of our method to PNG, we apply it to a test catalog of 125 FastPM-L1 simulations with $f_{\text{NL}} = 12$. These simulations were not used to calibrate the $f_{\text{NL}}$–moment relation, but lie within the interpolation range of the model ($f_{\text{NL}} \in [-25,25]$), making them ideal for validation.
        
        For this test, we restrict our analysis to a maximum scale of $s_{\mathrm{max}} = 275h^{-1}\,\mathrm{Mpc}$, chosen to optimize reliability given the simulation volume and sample size (see Appendix~\ref{app:sdep}). Accordingly, the expectation vector $\tilde{\mu}(f_{\text{NL}})$ (Eq.~\eqref{eqn:exp_vec}), observation vector $\hat{\mu}$ (Eq.~\eqref{eqn:obs_vec}), and covariance matrix $C_{ij}$ (Eq.~\eqref{eqn:cov}) are computed using 12 $s$-bins and the first three moments, yielding a vector of length $N = 36$ and a $36 \times 36$ covariance matrix.
        
        As a preliminary check, we show in Fig.~\ref{fig:quadfit} the moments from the averaged $f_{\text{NL}} = 12$ simulations (red stars) at several $s$-bins alongside the model's quadratic fit (dashed lines). The close agreement confirms the test case is well-described by our quadratic moment model, including even scales beyond $s_{\mathrm{max}}$ (shown for illustration).

\begin{figure}[htbp]\centering
        \includegraphics[width=9cm]{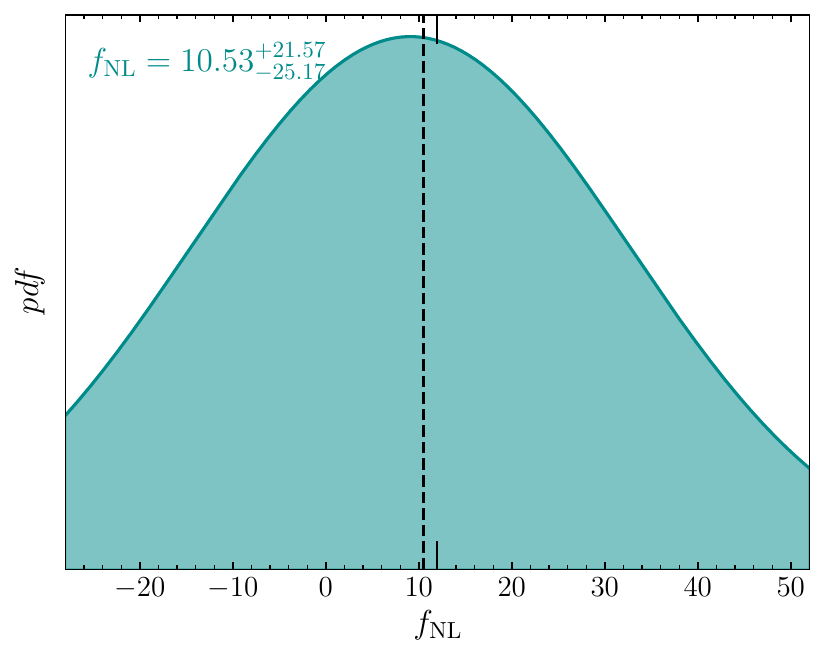}
        \caption{Gaussian fit of the marginalized posterior distribution on $f_{\text{NL}}$ from the MCMC sampling of the $f_{\text{NL}} = 12$ test case. The dashed line marks the median, with uncertainties given by the 16th and 84th percentiles (top-left). The true value $f_{\text{NL}} = 12$ is marked with extended ticks on the upper and lower $x$-axes.}
        \label{fig:test_all}
        \end{figure}

        We then sample the posterior using the affine-invariant Markov Chain Monte Carlo (MCMC) sampler \texttt{emcee} \citep{foreman-mackey2013}, based on Goodman and Weare’s algorithm \citep{goodman2010}. We use 30 walkers over 10,000 steps, discarding the first 1,000 as burn-in and thinning by half the autocorrelation time to retain independent samples. The resulting marginalized posterior on $f_{\text{NL}}$ is shown in Fig.~\ref{fig:test_all}, yielding $f_{\text{NL}} = 10.53^{+21.57}_{-25.17}$, consistent with the true value of $f_{\text{NL}} = 12$ within $0.1\sigma$ and exhibiting only a small bias of $\Delta f_{\text{NL}} = -1.47$, well within the expected uncertainty given the size of the mock ensemble.

        To further assess reliability, we compute the \emph{pull} statistic:
        \begin{equation}\label{eqn:pull}
            \mathrm{Pull} = \frac{f_{\text{NL},m} - f_{\text{NL}}^{\text{true}}}{\sigma_{f_{\text{NL},m}}},
        \end{equation}
        where $f_{\text{NL},m}$ and $\sigma_{f_{\text{NL},m}}$ are the measured value and 1$\sigma$ uncertainty from the $m$-th test simulation, and $f_{\text{NL}}^{\text{true}} = 12$.

        For each simulation, we compute $\hat{\mu}_{m,i}$ from the $m$-th box, derive $\tilde{\mu}_i(f_{\text{NL}})$ from the other 124 boxes, and compute the covariance matrix using the remaining 124 $f_{\text{NL}} = 0$ mocks. This ensures each pull measurement is fully independent of the model. A properly calibrated method should yield a pull distribution with mean zero and standard deviation one.

\begin{figure}[htbp]\centering
        \includegraphics[width=9cm]{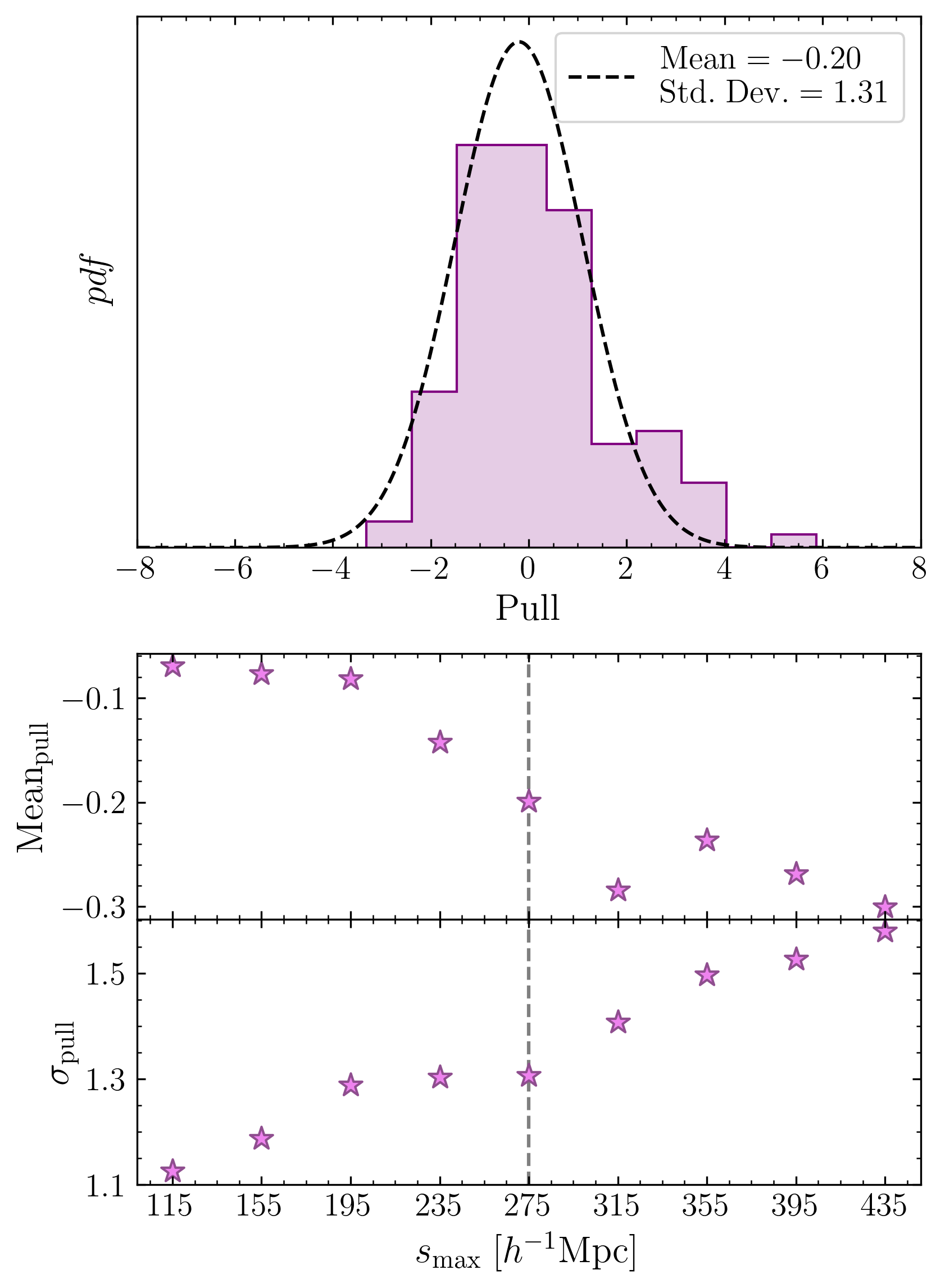}
        \caption{Top: distribution of the pull statistic (Eq.~\eqref{eqn:pull}) computed from the 125 $f_{\text{NL}} = 12$ simulations (purple). A Gaussian fit to the central region ($-3 < \text{Pull} < 3$; dashed line) yields a mean of $-0.20$ and standard deviation of $1.31$. Bottom: mean and standard deviation of the pull distribution as a function of $s_{\mathrm{max}}$, the largest radial bin included in the analysis. The vertical dashed line marks the adopted value $s_{\mathrm{max}} = 275h^{-1}\,\mathrm{Mpc}$ used in the test case.}
        \label{fig:pull}
        \end{figure}

        The resulting pull distribution is shown in Fig.~\ref{fig:pull}, with $\text{mean} = -0.20$ and $\text{std.\ dev.} = 1.31$. The slightly broadened distribution reflects the modest simulation count used for the $N = 36$-bin vector. We fit the pull histogram over the range $\text{Pull} \in [-3,3]$ to exclude a small number of non-Gaussian tails, which likely stem from sample noise (see Sect.~\ref{sec:disc:limits}).

    \subsection{Comparison of individual moments}\label{sec:mom}

        To better understand our model’s sensitivity to PNG and to assess how it compares to the standard two-point correlation function (2pcf), we analyze the contribution of each individual moment. Specifically, we decompose the full observation vector $\hat{\mu}_i$ into its components--$\mu_1(s)$, $\mu_2(s)$, and $\mu_3(s)$--and evaluate their respective constraining power on $f_{\text{NL}}$. We also examine the combined moment vector $\hat{\mu}_i = [\mu_1(s), \mu_2(s)]$ to test the impact of excluding the third moment.
        
\begin{figure}[htbp]\centering
        \includegraphics[width=9cm]{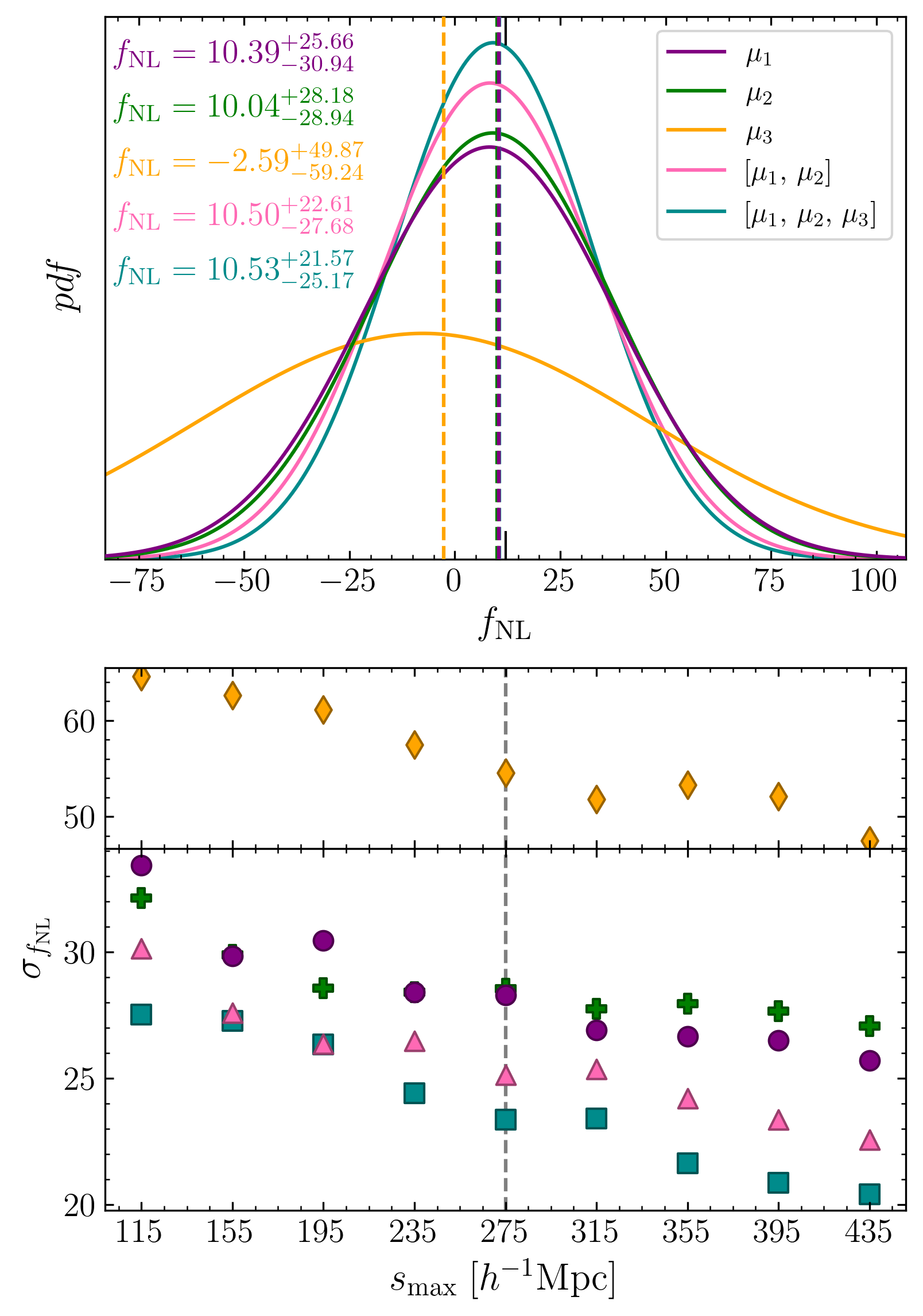}
        \caption{\textbf{Top panel:} Marginalized posterior distributions of $f_{\text{NL}}$ for five different observables: $\mu_1$ (purple), $\mu_2$ (green), $\mu_3$ (yellow), $[\mu_1,\, \mu_2]$ (pink), and $[\mu_1,\, \mu_2,\, \mu_3]$ (our full model, cyan; see Eq.~\eqref{eqn:obs_vec} and Fig.~\ref{fig:test_all}). Each observation vector is constructed and sampled independently using the $f_{\text{NL}}=12$ test simulations. Dashed lines indicate the median value of the MCMC posterior for each case, with corresponding uncertainties (16th and 84th percentiles) annotated in the upper left. \textbf{Bottom panel:} The $1\sigma$ uncertainty on $f_{\text{NL}}$ as a function of the maximum scale $s_{\mathrm{max}}$ used in the observation vector. Colors correspond to those in the top panel. Each point represents an independent sampling using bins up to that $s_{\mathrm{max}}$. The vertical gray dashed line marks the $s_{\mathrm{max}}$ adopted for the main analysis.}
        \label{fig:test_sep}
        \end{figure}

        For each moment combination, we independently fit the expectation vector coefficients $\mathcal{A}_i$, $\mathcal{B}_i$, and $\mathcal{C}_i$ and construct the corresponding statistical model as outlined in Sec.~\ref{sec:stat}. We then perform MCMC sampling using the same $f_{\text{NL}}=12$ simulation set described in Sec.~\ref{sec:test_case}. The resulting marginalized posteriors are shown in the top panel of Fig.~\ref{fig:test_sep}.

        The first and second moments perform similarly, each recovering $f_{\text{NL}}$ within $0.1\sigma$ of the true value. The third moment, $\mu_3(s)$, has significantly larger uncertainty and bias, recovering $f_{\text{NL}}$ within $0.3\sigma$. Nonetheless, including $\mu_3(s)$ in the full vector improves the overall constraint: adding the third moment reduces the uncertainty by $\sim8\%$ compared to the $[\mu_1,\,\mu_2]$ case and slightly lowers the bias. This improvement is consistent across a range of maximum scales, as shown in the bottom panel of Fig.~\ref{fig:test_sep} (pink points consistently below cyan).

        Combining the first two moments alone reduces the uncertainty by $\sim13\%$ compared to using either one individually. These results highlight the complementary information each moment provides and support the use of the full three-moment vector to maximize sensitivity to PNG.
        
        To contextualize these gains, we compare our method to the 2pcf. As defined in Eq.~\eqref{eqn:2pcf}, the 2pcf is equivalent to the spatial average of $\eta(\vec{x}, s)$, which corresponds to the first moment of the density perturbation estimator:
        \begin{equation}
            \mu_1(s)=\langle\Gamma_n(s)\rangle_n\propto\frac{\langle\eta(X,Y,Z;s)\rangle}{\langle \eta_R(X,Y,Z;s)\rangle} \sim \xi_2(s).
        \end{equation}
        That is, $\mu_1(s)$ captures the 2pcf signal normalized by the randoms correlation field $\langle \eta_R \rangle$. Therefore, comparing the full moments method ([$\mu_1,\, \mu_2,\, \mu_3$]) to $\mu_1$ alone is effectively a comparison of our moments method with the 2pcf. In this case, find a $\sim21\%$ reduction in the uncertainty on $f_{\text{NL}}$, indicating a significant sensitivity gain with minimal additional computational cost, since all three moments are derived from the same estimator.
        
\section{Discussion}\label{sec:discuss}

    Leveraging a suite of gigaparsec-scale halo simulations, we have demonstrated that a moments-based description of the density field recovers local primordial non-Gaussianity with percent-level accuracy and delivers $\sim 21\%$ tighter sensitivity to $f_{\mathrm{NL}}$ than the traditional two-point correlation function. Because sharpening $f_{\mathrm{NL}}$ constraints is essential for distinguishing single- from multi-field inflationary models—and thus for decoding the physics of the early Universe—the next challenge is to translate this proof-of-concept to real survey data. In what follows we therefore (i) lay out the practical steps needed to incorporate tracer bias and an optimal halo-mass cut, (ii) assess the numerical and volume-related limitations of our present simulation suite, and (iii) highlight the unexpectedly strong constraining power of the second density moment, which points the way to even greater gains.

\subsection{Tracer bias and mass cut}\label{sec:disc:mass}

    The statistical model introduced herein probes the cosmic matter density field for evidence of local PNG. Our method performs well in the validation tests described in this work. The primary goal of this study is to lay the groundwork for applying this methodology to observational data, particularly to large datasets with galaxy tracers such as those from DESI. However, several critical steps must be undertaken to adapt the model to such applications. Chief among these is the inclusion of the linear bias, $b_1$, of the tracer of interest. Since our method is simulation based, any tests performed using survey tracers must account for the difference in magnitude of the PNG signal between the halo simulations used to derive the model, and the observed tracer \citep[like the framework described in equations 15-19 of][]{Brown24}. While dark matter halos are a more direct tracer of the large-scale structure, galaxies introduce additional biases arising from selection effects, environmental influences, and population characteristics. Properly accounting for these effects is crucial to decouple the fiducial halo mass cut selection, $M_h$, from the model's predictions on observational data, thereby enhancing the model's applicability and robustness.
    
    A second clarification concerns our choice to omit \emph{a priori} PNG bias measurements and to adopt a mass cut that preserves the universality relation $p=1$.  In practice we measure the halo overdensity $\delta_{h}$, and the model ties the full expression—including any dependence on the PNG bias, $b_{\phi}$ or $p$—to the known input value of $f_{\mathrm{NL}}$ in the simulation catalogs.  Hence we are sensitive only to the degenerate product $f_{\mathrm{NL}}\,b_{\phi}$; the calibration absorbs the individual bias factors and renders them irrelevant for the present analysis.  We nevertheless impose a minimum halo mass, $M_{h,\mathrm{min}}$, chosen to keep $p=1$ in the halo mass function \citep{adame2024}.  This choice maximizes the separation of the $f_{\mathrm{NL}}\,b_{\phi}$ signal (via Eqs.~\eqref{eqn:NGden}–\eqref{eq:bphi}) and simplifies the treatment of the scale-dependent bias, reducing potential systematics and sharpening the eventual constraint on $f_{\mathrm{NL}}$.

\subsection{Limitations of our study}\label{sec:disc:limits}

    Our validation tests draw on four $L = 5h^{-1}\,\mathrm{Gpc}$ \textit{super}–simulations, each sliced into $125$ cubic sub–volumes of $L = 1h^{-1}\,\mathrm{Gpc}$. Treating those sub–boxes as independent allows us to assemble the $N_m = 125$ mocks used for the covariance matrix and moment uncertainties, but it implicitly assumes statistical homogeneity across the parent box. In real space that approximation appears largely benign: even a generous convolution kernel of radius $s = 500h^{-1}\,\mathrm{Mpc}$ still fits inside a sub–box, and edge effects are mitigated by (i) the integral–constraint correction built into \texttt{ConKer} \citep{Brown22} and (ii) the normalization by the mean random density contrast $\langle\eta_R(s)\rangle$ \citep[see also][]{riquelme2023}.  In Fourier space, however, each kernel is an average over \textit{all} wavevectors $\vec{k}$; long modes with $k \lesssim 0.006h\mathrm{Mpc}^{-1}$ (comparable to the fundamental mode of a $1h^{-1}\,\mathrm{Gpc}$ box) are shared by adjacent sub–boxes.  Consequently the same super–survey fluctuation feeds multiple "independent" realizations, and the number of \emph{truly} independent modes drops below unity for scales $s \gtrsim 180h^{-1}\,\mathrm{Mpc}$ (Eq.~\eqref{eqn:Nks}).  The result is a likely understatement of large–scale variance and an over-optimistic covariance matrix.  Crucially, this caveat reflects our mock construction, not an intrinsic flaw of the moments method; real surveys—or mock suites generated as separate boxes—naturally include the full super-sample covariance matrix.

    A related issue is the size of the mock ensemble. The fiducial covariance matrix is $36 \times 36$ but is derived from only $N_m = 125$ mocks, whereas the rule of thumb $N_m \gtrsim 10N$ would suggest $N_m \gtrsim 360$. The widened pull distribution ($\sigma_{\mathrm{pull}} = 1.31$; Fig.~\ref{fig:pull}) indicates that statistical noise inflates our error bars. Generating an order‐of‐magnitude larger ensemble is straightforward—modern surveys routinely provide thousands of low-fidelity mocks for precisely this purpose.

    A third systematic concerns the scale range and binning. Our sensitivity to $f_{\mathrm{NL}}$ saturates at the largest kernel radii probed, $s_{\max}$, which are capped by the catalogue volume (Appendix~\ref{app:sdep}). Kernels approaching the full $(1h^{-1}\,\mathrm{Gpc})^{3}$ box become noisy and unreliable. Larger‐volume data sets such as the \textit{DESI}~DR2 catalog $(2.4h^{-1}\,\mathrm{Gpc})^{3}$ would admit longer modes, tighten the constraints (see the decreasing trend with $s_{\max}$ in Fig.~\ref{fig:test_sep}), and support finer binning (e.g. $\Delta s = 4h^{-1}\,\mathrm{Mpc}$ in \citealt{DESI2025}, versus our current $20h^{-1}\,\mathrm{Mpc}$).

    Lastly, we revisit our working assumption that the moment vector depends \emph{quadratically} on $f_{\mathrm{NL}}$. Sec.~\ref{sec:den_moments} shows that each individual moment scales as $\mu_n \propto f_{\mathrm{NL}}^{2n}$, so only $\mu_1$ is strictly quadratic; higher moments contain higher-order terms that we truncate. This approximation holds up both visually--each moment adheres closely to a quadratic fit across scales (Fig.~\ref{fig:quadfit})--and empirically, as the full moment vector delivers strong constraining power (Fig.~\ref{fig:test_all}). Residuals become noticeable only for $s > 300h^{-1}\,\mathrm{Mpc}$, where the $f_{\mathrm{NL}} = 12$ test begins to deviate in $\mu_3$. While a denser simulation grid in $f_{\mathrm{NL}}$ could in principle model these higher-order terms, the deviation remains well within our error bars, and both LSS and CMB data already place $f_{\mathrm{NL}}^{\mathrm{loc}}$ comfortably in the $[-10, 10]$ range. For these reasons, the quadratic truncation is more than sufficient for our analysis. Future work may explore higher-order models for completeness, though we do not expect a meaningful gain in sensitivity.

    In short, the dominant systematics--box coupling, limited mock statistics, and volume-driven scale cuts--stem from our testbed rather than from the method itself. All can be alleviated with larger‐volume, higher-multiplicity mock catalogues, and exploring those upgrades is a clear next step.

\subsection{Promise of the second moment}

    Beyond addressing these limitations, our analysis uncovers an unexpected strength: the second moment alone carries nearly as much constraining power as the mean (see Fig.~\ref{fig:test_sep}). As we showed in Sec.~\ref{sec:mom}, $\mu_1\sim\xi_2$, so comparing the first and second moments is like comparing the familiar 2pcf (the average of the density perturbations) to $\mu_2$ (the variance of the density perturbations). Since both moment vectors contain the same number of bins, this comparison is not confounded by binning resolution or covariance matrix smoothness. Fig.~\ref{fig:test_sep} confirms that $\mu_2$ offers comparable sensitivity to $f_{\mathrm{NL}}$ as $\mu_1$, while Fig.~\ref{fig:moments} further shows that $\mu_2$ more clearly tracks the scale-dependent bias, evident in the growing separation between different $f_{\mathrm{NL}}$ cases in the lower center panel. This pattern is also visible in the density field distributions (Fig.~\ref{fig:distributions}), where the mean remains similar across scales, but the width (i.e., $\mu_2$) narrows significantly. Finally, the physical interpretation of $\mu_2$ given in Appendix~\ref{app:momphys} suggests that the second (and higher) moment(s) may surpass $\mu_1$ in sensitivity in larger-volume surveys. Altogether, these findings point to a promising path: analyses of PNG in the LSS may benefit from combining perturbation moments (or even using the variance alone)--rather than relying solely on mean-based statistics like the two-point function. This approach is computationally efficient and potentially more sensitive to PNG than the classic 2pcf.

\subsection*{Outlook}

    Looking forward, three incremental yet concrete steps will carry this proof-of-concept across the finish line to a survey-ready analysis. \textit{First}, explicitly embedding the tracer’s linear \emph{and} scale-dependent bias in our forward model—together with a well-motivated lower halo-mass threshold--will disentangle the $f_{\mathrm{NL}}$ imprint from galaxy-selection systematics and extend our calibration beyond dark matter halos. \textit{Second}, deriving the covariance matrix from a mock ensemble at least an order of magnitude larger than the present $N_m=125$ will quench statistical noise, stabilize large-scale bins, and permit finer radial binning. \textit{Third}, deploying the method on volumes of order $(2.4h^{-1}\,\mathrm{Gpc})^{3}$—for example, those available in DESI~DR2—will admit modes with $k\lesssim0.003h\,\mathrm{Mpc}^{-1}$ and further amplify the constraining power, especially in the second and higher moments. Work along these lines is already under way and will be detailed in a forthcoming paper.

\section{Conclusions}

    In this work, we introduced a new framework for probing primordial non-Gaussianity using the moments of the density perturbation field. By computing the first three moments of the normalized two-point estimator $\Gamma_n(s)$ from dark matter halo simulations seeded with varying $f_{\text{NL}}$, we built a statistical model that captures the scale-dependent signatures of local PNG. Applying this model to an independent $f_{\text{NL}}=12$ test simulation set, we demonstrated percent-level recovery and a $\sim21\%$ improvement in sensitivity over the traditional two-point correlation function.
    
    These results highlight the potential of moments-based observables as efficient and powerful summary statistics for LSS analyses. In particular, the second moment alone was shown to carry nearly equivalent constraining power to the mean, suggesting that higher moments can extract additional PNG signal without significant added computational cost.
    
    Future refinements--including improved treatment of tracer bias, expanded simulation ensembles, and modeling higher-order dependencies in $f_{\text{NL}}$--will be essential for adapting this method to observational data. As ongoing surveys like DESI deliver increasingly large and precise catalogs, moments-based analyses offer a scalable path to sharpening constraints on primordial physics and advancing our understanding of the inflationary universe.

\appendix

\section{Physical interpretation of higher moments}\label{app:momphys}
    
    In Sec.~\ref{sec:den_moments}, we introduced the empirical definitions of the Gaussian moments (Eqs.~\eqref{eqn:mu1}–\eqref{eqn:mu3}). Here, we derive an analytic expression for the second moment, $\mu_2(s)$, to clarify its physical meaning and justify its value in constraining PNG. We omit $\mu_1$ since it corresponds to the usual 2pcf, and $\mu_3$ since its interpretation follows naturally from this treatment.
    
    The second moment--i.e., the variance--is given by
    \begin{align}
        \mu_2(s) &= \langle (\delta(x)\delta(x+s) - \mu_1(s))^2 \rangle \\
        &= \langle \delta^2(x)\delta^2(x+s) \rangle - \mu_1^2(s).
    \end{align}
    The first term encodes the connected four-point function and thus contains the non-Gaussian signal. For simplicity, we omit the randoms normalization in the above expression, as it does not affect the present discussion. Its Fourier transform is
    \begin{gather}\label{eqn:mu2_fourier}
        \langle \delta^2(x)\delta^2(x+s) \rangle = \iiiint 
        \frac{d^3\vec{k}_1\,d^3\vec{k}_2\,d^3\vec{k}_3\,d^3\vec{k}_4}{(2\pi)^{12}} \nonumber \\
        \times \langle \delta(\vec{k}_1)\delta(\vec{k}_2)\delta(\vec{k}_3)\delta(\vec{k}_4) \rangle \, e^{i\vec{x}\cdot\vec{k}_{1234}} e^{i\vec{s}\cdot(\vec{k}_2+\vec{k}_4)},
    \end{gather}
    with $\vec{k}_{1234} \equiv \vec{k}_1+\vec{k}_2+\vec{k}_3+\vec{k}_4$. This is a collapsed limit of the 4-point function, with two positions at $x$ and two at $x+s$. While $s$ appears as a vector in the exponential, we average over all directions on a sphere of radius $|\vec{s}|$ to recover the scalar separation $s$ used in the moments.
    
    Applying Wick’s theorem, the 4-point expectation splits into Gaussian and connected parts:
    \begin{gather}
        \langle \delta(\vec{k}_1)\delta(\vec{k}_2)\delta(\vec{k}_3)\delta(\vec{k}_4) \rangle = 
        [\langle \delta(\vec{k}_1)\delta(\vec{k}_2) \rangle \langle \delta(\vec{k}_3)\delta(\vec{k}_4) \rangle \nonumber \\
        + 2\,\text{permutations}] + \langle \delta(\vec{k}_1)\delta(\vec{k}_2)\delta(\vec{k}_3)\delta(\vec{k}_4) \rangle_{\text{c}}.
    \end{gather}
    The Gaussian contractions are already captured in $\mu_1^2$. The connected component contributes:
    \begin{equation}
        \langle \delta(\vec{k}_1)\delta(\vec{k}_2)\delta(\vec{k}_3)\delta(\vec{k}_4) \rangle_{\text{c}} = 
        (2\pi)^3 \delta_D(\vec{k}_{1234})\, \mathcal{T}(k_1, k_2, k_3, k_4),
    \end{equation}
    where $\mathcal{T}$ is the trispectrum. The Dirac delta enforces momentum conservation and reduces the integral to three independent $k$-vectors:
    \begin{gather}
        \langle \delta^2(x)\delta^2(x+s) \rangle_{\text{c}} = \iiiint \frac{d^3\vec{k}_1\,d^3\vec{k}_2\,d^3\vec{k}_3\,d^3\vec{k}_4}{(2\pi)^{12}}\, 
        (2\pi)^3 \nonumber \\
        \hspace{2em}\times \delta_D(\vec{k}_{1234})\mathcal{T}(k_1, k_2, k_3, k_4)e^{i\vec{x}\cdot\vec{k}_{1234}} e^{i\vec{s}\cdot(\vec{k}_2+\vec{k}_4)} \\
        = \iiint \frac{d^3\vec{k}_1\,d^3\vec{k}_2\,d^3\vec{k}_3}{(2\pi)^6}\, 
        \mathcal{T}(k_1, k_2, k_3, -k_{123})\, e^{-i\vec{s}\cdot(\vec{k}_1 + \vec{k}_3)}.
    \end{gather}
    This expression emphasizes collapsed configurations ($\vec{k}_1 + \vec{k}_3 \approx 0$), since the real-space correlator involves only two distinct points. These configurations are especially sensitive to local PNG, which produces long-short mode coupling. The full 4-point function $\langle \delta(x_1)\delta(x_2)\delta(x_3)\delta(x_4)\rangle_c$ would span a wider set of shapes (e.g., equilateral, orthogonal), but $\mu_2$ selectively probes the collapsed limit relevant to local-type non-Gaussianity.
    
\subsection*{Survey volume sensitivity}
    
    This physical picture offers insight into how moment sensitivity scales with volume. In a box of side length $L$ and volume $V = L^3$, the fundamental mode is
    \begin{equation}
    \vec{k}_f = \frac{2\pi}{L} \vec{n}, \quad \vec{n} \in \mathbb{Z}^3.
    \end{equation}
    The number of $k$-modes in a spherical shell of radius $k$ and width $\Delta k$ is:
    \begin{equation}
    N_k(k) = \frac{4\pi k^2 \Delta k}{(2\pi/L)^3} = \frac{k^2 \Delta k}{2\pi^2} V.
    \end{equation}
    Switching to real-space scale $s \sim 1/k$ with $\Delta k \approx \Delta s/s^2$ yields:
    \begin{equation}
    N_k(s) \approx \frac{\Delta s}{2\pi^2 s^4} V. \label{eqn:Nks}
    \end{equation}
    So the number of available $k$-modes falls sharply with increasing $s$. For example, with $V = 1\,h^{-3}\,\mathrm{Gpc}^3$, $s = 55h^{-1}\,\mathrm{Mpc}$, and $\Delta s = 20h^{-1}\,\mathrm{Mpc}$, we find $N_k \approx 111$--and fewer at larger scales.
    
    Now, since $\mu_1$ samples a single $k$-vector and $\mu_2$ integrates over three wavevectors (constrained by $\vec{k}_{1234}=0$), the number of available mode combinations scales as:
    \begin{align}
    N_k^{\mu_1} &\sim N_k, \\
    N_k^{\mu_2} &\sim N_k^3.
    \end{align}
    While the Dirac delta imposes constraints, $\mu_2$ still integrates over a far larger space of configurations than $\mu_1$. This explains its enhanced constraining power in finite-volume surveys and highlights the promise of higher moments for probing PNG. Furthermore, since $N_k\propto V$, increases in survey volume significantly enhance the relative signal of higher moments.

\section{Dependence of the covariance matrix on $f_{\text{NL}}$}\label{app:cov}

\begin{figure}[htbp]\centering
    \includegraphics[width=9cm]{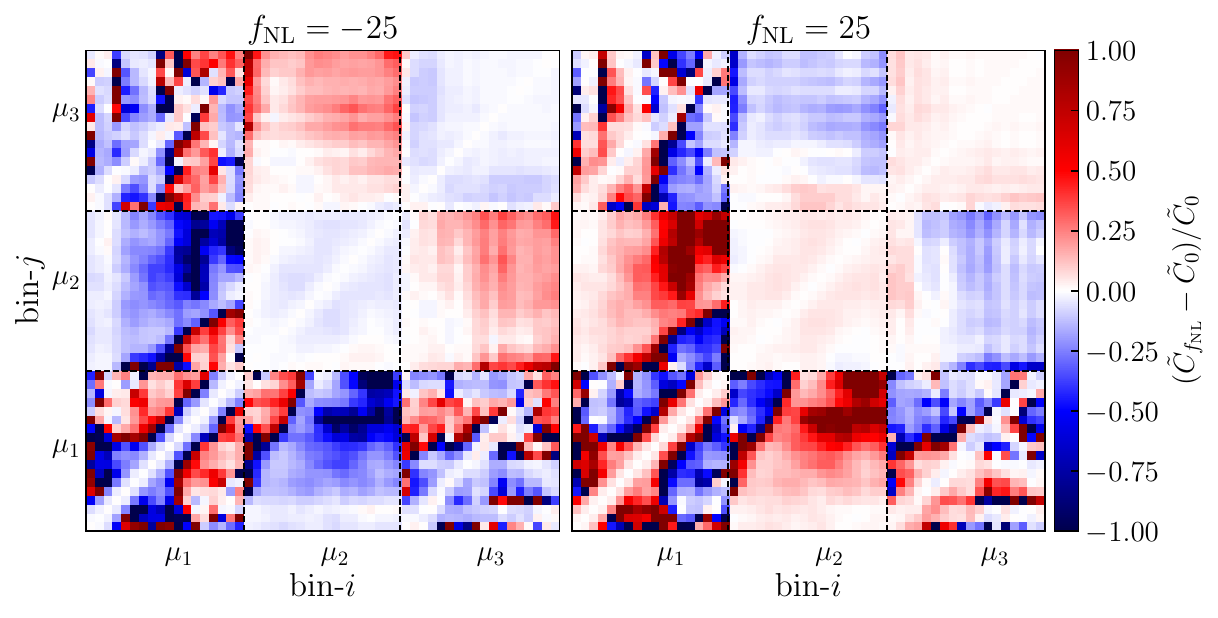}
    \caption{The relative difference between the correlation matrix ($\tilde{C_{ij}}=C_{ij}/\sqrt{C_{ii}C_{jj}}$) derived from the $f_{\text{NL}} = -25$ (left) and $f_{\text{NL}} = 25$ (right) simulations, and that from the $f_{\text{NL}} = 0$ fiducial simulations.}
    \label{fig:covdiff}
    \end{figure}

    In this study, we compute the covariance matrix using the fiducial $f_{\text{NL}}=0$ FastPM-L1 simulations, following standard practice in similar analyses \citep[e.g.,][]{mueller19,mueller21,DESI24,Rezaie21}. To assess the robustness of this assumption, we compare the fiducial covariance matrix with those derived from the $f_{\text{NL}} = -25$ and $f_{\text{NL}} = 25$ FastPM-L1 simulations. The relative differences in the normalized covariance matrices are shown in Fig.~\ref{fig:covdiff}.
    
    We observe statistically significant deviations between the covariance matrices. In particular, relative differences reach $\mathcal{O}(1)$ in some regions and average around $\mathcal{O}(0.1)$ across the matrix. Notably, the $f_{\text{NL}} = -25$ and $f_{\mathrm{NL}}=25$ covariance matrices seem to diverge from the fiducial case in opposite directions (illustrated by the nearly inverted colors across the 0-point in Fig.~\ref{fig:covdiff}).
    
    These results suggest a non-negligible dependence of the covariance matrix structure on the sign and magnitude of the primordial non-Gaussianity with which the simulations are seeded. Since our method probes PNG directly, this dependence suggests a potentially similar impact on other PNG-sensitive observables, such as higher-order $n$-point correlation functions or the bispectrum. Future studies might explore alternatives such as deriving the covariance matrix from $\ln P_m$, where $P_m$ is the matter power spectrum, which has been shown to exhibit approximate independence from $f_{\text{NL}}$ \citep[see Eq.~13 of][]{Ross13}.

\section{Scale dependence}\label{app:sdep}

    Here we analyze the scale dependence of our model and justify the choice of $s_{\mathrm{max}} = 275h^{-1}\,\mathrm{Mpc}$ and bin size $\Delta s = 20h^{-1}\,\mathrm{Mpc}$ for constructing the observation vector. While PNG signals grow stronger on large scales due to long-short mode coupling, our simulation boxes span only $L = 1h^{-1}\,\mathrm{Gpc}$, so spherical kernels with $s \gtrsim 500h^{-1}\,\mathrm{Mpc}$ intersect box boundaries and yield noisy measurements. To avoid this, we limit our tests to $s_{\mathrm{max}} \leq 435h^{-1}\,\mathrm{Mpc}$, maintaining a conservative buffer from the edges.

    To determine the optimal $s_{\mathrm{max}}$, we run our statistical pipeline at nine values between $115$ and $435h^{-1}\,\mathrm{Mpc}$, measuring $f_{\mathrm{NL}}$ and its uncertainty at each. As shown in the bottom panel of Fig.~\ref{fig:test_sep}, larger $s_{\mathrm{max}}$ yields tighter constraints, as expected. However, increasing the number of bins also enlarges the covariance matrix, worsening its condition given the limited number of simulations (see Sec.~\ref{sec:disc:limits}).

    To quantify this trade-off, we perform a pull analysis at each $s_{\mathrm{max}}$ (bottom panel of Fig.~\ref{fig:pull}). We find that pulls become increasingly biased and their uncertainties increasingly underestimated at large scales: $\mathrm{Mean}_{\mathrm{pull}}$ drifts negative, and $\sigma_{\mathrm{pull}} > 1$. As shown in the bottom panel of Fig.~\ref{fig:pull}, both metrics are stable in the range $195$–$275h^{-1}\,\mathrm{Mpc}$: the bias remains moderate ($\mathrm{Mean}_{\mathrm{pull}} \approx -0.1$ to $-0.2$), and the spread stays relatively stable ($\sigma_{\mathrm{pull}} \approx 1.3$). Beyond $s_{\mathrm{max}} = 275h^{-1}\,\mathrm{Mpc}$, the pull distribution becomes increasingly skewed and its variance grows rapidly. This plateau identifies $275h^{-1}\,\mathrm{Mpc}$ as the upper limit beyond which instability dominates. We therefore adopt $s_{\mathrm{max}} = 275h^{-1}\,\mathrm{Mpc}$ for testing, as it balances information gain from large-scale modes with robustness of the statistical estimator.

    The bin width $\Delta s = 20h^{-1}\,\mathrm{Mpc}$ (yielding 36 bins) is chosen similarly. Finer binning increases vector dimensionality and amplifies noise in the covariance matrix, degrading diagnostics like the pull test. We find that moderate variations around $\Delta s = 20$ have negligible impact on PNG sensitivity but reduce numerical stability. Thus, the chosen pair $(s_{\mathrm{max}},\,\Delta s) = (275h^{-1}\,\mathrm{Mpc},\,20h^{-1}\,\mathrm{Mpc})$ reflects a practical minimum in statistical uncertainty and numerical noise given our simulation suite.
 
\begin{acknowledgements}
    E.V.’s stipend at SAO is supported through the Chandra X-Ray Center grant GO3-24093X and Space Telescope Science Institute grant HST-GO-17177. Z.B. is grateful for support from the US Department of Energy via grants DE-SC0021165 and DE-SC0011840. R.D. acknowledges support from the Department of Energy under the grant DE-SC0008475.0. E.C. is supported by the Director, Office of Science, Office of High Energy Physics of the U.S. Department of Energy under Contract No. DE-AC02-05CH1123, and by the National Energy Research Scientific Computing Center, a DOE Office of Science User Facility under the same contract.
\end{acknowledgements}

\bibliographystyle{aa} 
\bibliography{myreferences}

\end{document}